\documentclass[runningheads]{llncs}

\usepackage[hyphens]{url}
\usepackage[hidelinks]{hyperref}
\usepackage[utf8]{inputenc}
\usepackage{subcaption}
\usepackage{mathabx}
\usepackage{microtype}
\usepackage{multirow}

\newcommand\Small{\fontsize{9}{9.2}\selectfont}
\newcommand*\LSTfont{%
	\Small\ttfamily\SetTracking{encoding=*}{-60}\lsstyle}

\usepackage{listings}
\usepackage{tikz}

\begin{document}

\title{Data Preparation in Agriculture Through Automated Semantic Annotation – Basis for a Wide Range of Smart Services}
\titlerunning{Data Preparation in Agriculture Through Automated Semantic Annotation}
\authorrunning{Klose et al.}

\author{
	Julian Klose\inst{1} \and
	Markus Schröder\inst{2} \and
	Silke Becker\inst{1} \and
	Ansgar Bernardi\inst{2} \and
	Arno Ruckelshausen\inst{1}
}
\institute{
	Osnabrück University of Applied Sciences, Faculty IuI, Osnabrück, Germany\\
	\email{\{julian.klose, s.becker.1, a.ruckelshausen\}@hs-osnabrueck.de},
	\and
	Smart Data \& Knowledge Services Dept., DFKI GmbH, Kaiserslautern, Germany\\ \and
	Computer Science Dept., TU Kaiserslautern, Germany\\
	\email{\{markus.schroeder, ansgar.bernardi\}@dfki.de},
}

%

\maketitle

\begin{abstract}

Modern agricultural technology and the increasing digitalisation of such processes provide a wide range of data. However, their efficient and beneficial use suffers from legitimate concerns about data sovereignty and control, format inconsistencies and different interpretations. As a proposed solution, we present Wikinormia, a collaborative platform in which interested participants can describe and discuss their own new data formats. Once a finalized vocabulary has been created, specific parsers can semantically process the raw data into three basic representations: spatial information, time series and semantic facts (agricultural knowledge graph). Thanks to publicly accessible definitions and descriptions, developers can easily gain an overview of the concepts that are relevant to them. A variety of services will then (subject to individual access rights) be able to query their data simply via a query interface and retrieve results. We have already implemented this proposed solution in a prototype in the SDSD (Smart Data - Smart Services) project and demonstrate the benefits with a range of representative services. This provides an efficient system for the cooperative, flexible digitalisation of agricultural workflows.

\end{abstract}

\section{Introduction}

Agriculture is confronted with constantly growing requirements such as documentation, traceability, process optimisation and resource-saving work. At the same time, there is a constant development in the quantitative and qualitative performance of agricultural machinery that leads to increasing availability of data, which is usually mapped manufacturer-specifically or machine-specifically. The cross-manufacturer data exchange and the correct interpretation and use of various data formats and contents are the farmers' major challenges - especially in view of dynamic advancements. 

In order to meet these requirements, we have developed a system in the research project Smart Data - Smart Services\footnote{\url{http://www.sdsd-projekt.de/}} (SDSD) that combines enriched agricultural process data with intelligent services. Our approach stores information from different agricultural sources of a farmer. By establishing storage and access rules, the user retains data sovereignty and control.

A major challenge lies in the recognition and interpretation of the different data formats and contents. Standardization often takes too long to adapt dynamically to new requirements. SDSD offers a platform for the collaborative description of data formats, which we call Wikinormia. This open platform enables us to achieve a high level of extensibility that keeps pace with the ongoing development of new data formats. The use of Semantic Web Standards makes it possible to link the modelled information to a knowledge graph.

In addition to data preparation, SDSD is also a platform for providing services that generate benefits from smart data, such as automated documentation, calculations and application aids, or long-term analyses. Services can retrieve the required information from the linked data without coming into contact with the different formats of the raw data. This system shows first successes for a beneficial digitalisation of agricultural processes.

\section{Related Work}
\label{sec:relwork}

There are already partial solutions and related research projects in this topic. Industrial approaches such as MyJohnDeere\footnote{\url{https://myjohndeere.deere.com/}} and 365FarmNet\footnote{\url{https://www.365farmnet.com/}} are platforms for transmitting and storing machine data and task data. However, these platforms are purely proprietary and only compatible with the respective machines. With the diverse use of machines from different manufacturers, a manufacturer-independent system is unavoidable.

DKE's agrirouter\footnote{\url{https://my-agrirouter.com/}} provides an open data transfer platform that allows task data and real-time telemetry data to be transferred from the machines to connected applications and vice versa. However, the data packets are not opened by the agrirouter and remain stored only for transmission purposes for a short time. With DKE as our partner in SDSD, we have successfully used the agrirouter to simplify communication to and from the SDSD system.

Other research projects, such as GeoBox\footnote{\url{https://www.dfki.de/web/forschung/projekte-publikationen/projekte/projekt/geobox/}}, ODiL\footnote{\url{https://saat.dfki.de/de/projekte/odil.html}} and OPeRAte\footnote{\url{http://operate.edvsz.hs-osnabrueck.de/}}, are working on similar topics for agricultural data management and supply. SDSD differs from these projects, among other things, by the open extensible format description using Wikinormia. Furthermore, SDSD cooperates with the recently launched ATLAS\footnote{\url{https://cordis.europa.eu/project/rcn/223982/factsheet/en}} project.

\section{Approach}
\label{sec:approach}

\begin{figure}[h]
	\centering
	\includegraphics[width=1.0\textwidth]{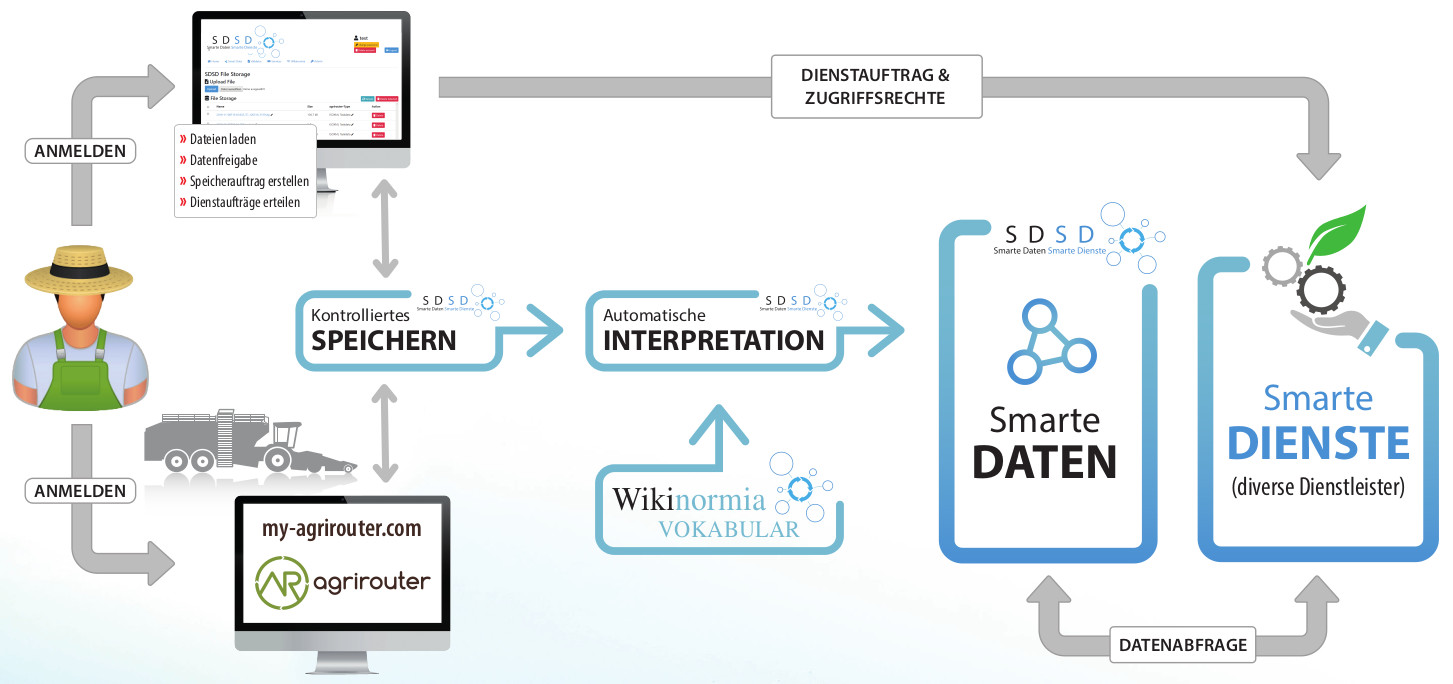}
	\caption{Overview of the SDSD components (in German): Via agrirouter or our web interface data can be stored, which is automatically interpreted by the Wikinormia vocabulary to create smart data. Smart services query the processed data.}
	\label{diagram}
\end{figure}
\begin{figure}[h]
	\centering
	\includegraphics[width=1.0\textwidth]{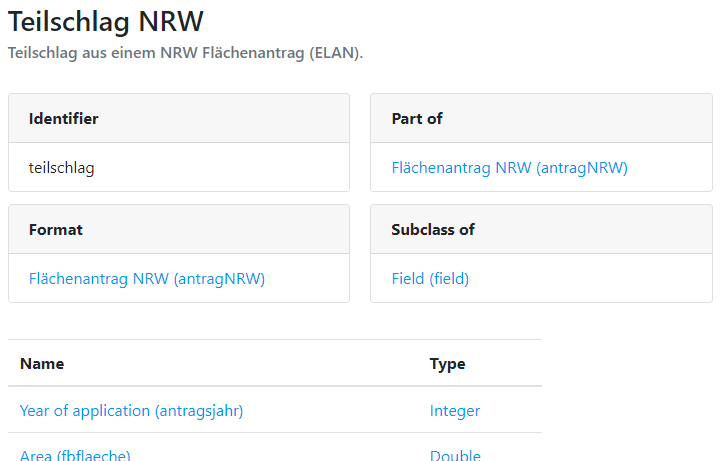}
	\caption{Wikinormia entry for North Rhine-Westphalia (NRW) agricultural application (partly in German).}
	\label{wikinormia}
\end{figure}

\noindent
Our main approach (\autoref{diagram}) is the generation of smart data and its provision for smart services. In order to convert the received raw data into smart data, the format must first be described. SDSD offers the Wikinormia as a Wiki-like platform for this purpose: A format is defined by the Resource Description Framework\cite{rdf} (RDF) in the form of a graph. This definition should be made by the data providers. For publicly accessible formats, this can also be done by service developers. This is not intended to replace standardization, but rather to significantly accelerate the development process. Our prototype Wikinormia already describes the widely used ISOXML\cite{aef} standard, but also more specific data formats such as North Rhine-Westphalia (NRW) agricultural application data (see \autoref{wikinormia}).

When inserting a new data format in Wikinormia, a parser must also be provided, which can read the data format from the raw data. This is necessary due to the large number of possible transmission formats (e.g. JSON, XML, binary). In special cases a parser could be generated automatically based on the Wikinormia description.

\newpage
\noindent
SDSD distinguishes between three data models and uses specialized databases (polyglot storage\cite{fowler2012introduction}):
\begin{enumerate}
	\item Spatial data includes positioned geometries, such as lanes or field boundaries. They are stored in a MongoDB\footnote{\url{https://www.mongodb.com}} with spatial indexing. This makes it possible to search for overlaps or elements in spatial proximity.
	\item Time series contain time-related and position-related sensor data, such as machine speeds or specific application rates. Due to their quantity, they are stored in a Cassandra database\footnote{\url{https://cassandra.apache.org/}} that is optimized for storing large tables with variable column lengths.
	\item Semantic graphs contain linked information, such as machine descriptions or tasks. They are saved in a Stardog Triplestore\footnote{\url{https://www.stardog.com/}}. This allows that relationships between information can be queried in a structured way, such as "all task in which a sowing machine was used".
\end{enumerate}
Instances (machines, tasks, fields, persons, etc.) receive unique addresses (URIs) in the RDF graph. Thus, it becomes possible to connect geodata and time series with the instances and to link them with the corresponding description in the Wikinormia. Linking makes it easy to annotate additional information directly to an instance, e.g. weather information to a field. Deduplication of the same instances from different sources is also simplified. If, for example, a field that is already known from an area application occurs again in an task documentation (ISOXML TaskData), the system can link the two field instances with a \textit{SameAs} relationship. The equality is based on spatial information.

The generated smart data can be retrieved by smart services. Developers of smart services no longer have to deal with the integration of different data formats or the peculiarities of individual data sources. Selected information required by the service can be retrieved via the SDSD interface. This enables even small providers to deploy their services cost-effectively. In addition, the farmer can specify exactly what information the service is allowed to access and thus protect confidential data.
\begin{figure}[h]
	\centering
	\includegraphics[width=1.0\textwidth]{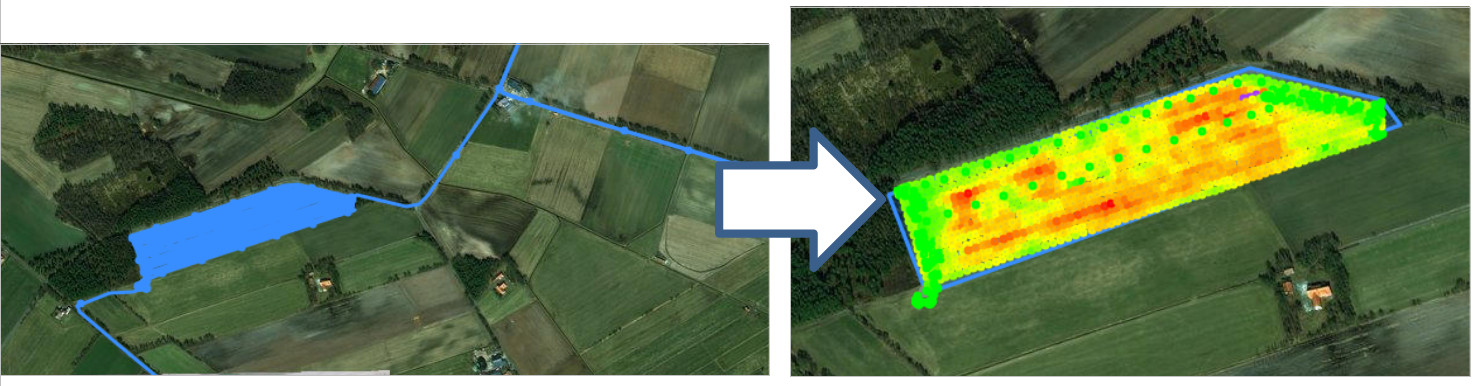}
	\caption{Time series separation service: Left a contiguous time series, right an automatically separated field with yield mapping.}
	\label{application}
\end{figure}
The time series separation service in \autoref{application} solves a typical problem with recorded machine data: In addition to the actual field, road trips and the processing of other fields are also stored in the task data. Such recording occurs when tasks are not properly started and stopped by the driver. The example service reads such data (left) and overlays it with already known field boundaries to separate the relevant information (right). If the fields are not yet known, field boundaries from OpenStreetMap\footnote{\url{https://wiki.openstreetmap.org/wiki/Overpass_API}} are taken into account. This way, SDSD combines field information from various sources and makes it available for retrieval via spatial query.

A prototype for testing SDSD is available online\footnote{\url{https://app.sdsd-projekt.de}} with username "sdsd" and password "sdsd".

\section{Conclusion and Outlook}
\label{sec:concl}

In order to map the processes in agriculture digitally, we need a joint approach by all those involved. SDSD provides a platform for communication between data producers, farmers and service providers. 

The SDSD platform described above can be a solution for the upcoming challenges of digitalisation in agriculture. This includes a) the description of data formats in the Wikinormia, b) the readout by parsers, c) the semantic processing and d) the controlled provision of information for services.

Through the targeted use and integration of data, short and medium-term management strategies are optimised. This can have a positive effect on process chains and the use of resources. In the long run, the farmer builds a cultivation history by using the SDSD platform. 


\bibliographystyle{llncs}
\bibliography{paper}

\begin{thebibliography}{1}
\providecommand{\url}[1]{\texttt{#1}}
\providecommand{\urlprefix}{URL }
\providecommand{\doi}[1]{https://doi.org/#1}

\bibitem{aef}
{AEF}: {ISO 11783} (2015)

\bibitem{fowler2012introduction}
Fowler, M., Sadalage, P.J.: Introduction to polyglot persistence: Using
  different data storage technologies for varying data storage needs  (2012)

\bibitem{rdf}
{W3C}: {Resource Description Framework} (2014),
  \url{https://www.w3.org/TR/rdf11-primer/}

\end{thebibliography}

\end{document}